**TITLE PAGE** **Original**

(1) PCP-3/ISS 2008

(2) Submitted: November 4 2008

(3) Title:

Synthesis of infinite-layer $LaNiO_2$ films by metal-organic deposition

(4) Author's Names

D. Kaneko[a], K. Yamagishi[a], A. Tsukada[b], T. Manabe[c], M. Naito[a]

(5) Addresses

[a] Department of Applied Physics, Tokyo University of Agriculture and Technology, Naka-cho 2-24-16, Koganei, Tokyo 184-8588, Japan

[b] Geballe Laboratory for Advanced Materials, Stanford University, Stanford, California 94305, USA

[c] National Institute of Advanced Industrial Science and Technology (AIST), Higashi 1-1-1, Tsukuba, Ibaraki 305-8565, Japan

*) Corresponding author. Address: Department of Applied Physics, Tokyo University of Agriculture and Technology, Naka-cho 2-24-16, Koganei, Tokyo 184-8588, Japan. Tel. +81 42 388 7229; fax: +81 42 385 6255. E-mail address: minaito@cc.tuat.ac.jp.

PCP-3/ISS 2008

# Synthesis of infinite-layer $LaNiO_2$ films by metal organic decomposition

D. Kaneko[a], K. Yamagishi[a], A. Tsukada[b], T. Manabe[c], and *M. Naito[a]

[a] Department of Applied Physics, Tokyo University of Agriculture and Technology,

Naka-cho 2-24-16, Koganei, Tokyo 184-8588, Japan

[b] Geballe Laboratory for Advanced Materials, Stanford University,

Stanford, California 94305, USA

[c] National Institute of Advanced Industrial Science and Technology (AIST),

Higashi 1-1-1, Tsukuba, Ibaraki 305-8565, Japan

**Abstract**

We report the synthesis of infinite-layer $LaNiO_2$ thin films by metal organic decomposition. Our work is aimed to synthesize perovskite-like oxides with $3d^9$ electronic configuration, which is common to high-$T_c$ copper oxides. The $3d^9$ configuration is very rare in oxides other than cuprates. $Ni^{1+}$ oxides, even though $Ni^{1+}$ is an unusual oxidation state, may be one of very few candidates. One example of the $Ni^{1+}$ phases is infinite-layer $LaNiO_2$. The bulk synthesis of $LaNiO_2$ is difficult, but we demonstrate in this article that the thin-film synthesis of $LaNiO_2$ by metal organic decomposition is rather easy. This is due to the advantage of thin films with a large-surface-to-volume ratio, which makes oxygen diffusion prompt. Resistivity measurements indicate that $LaNiO_2$ is essentially a metal but unfortunately with no trace of superconductivity yet.

*) Corresponding author. Address: Department of Applied Physics, Tokyo University of Agriculture and Technology, Naka-cho 2-24-16, Koganei, Tokyo 184-8588, Japan. Tel. +81 42 388 7229; fax: +81 42 385 6255. E-mail address: minaito@cc.tuat.ac.jp.

## 1. Introduction

More than two decades have passed since the discovery of high-temperature superconductivity in cuprates, and many (more than 100) superconducting cuprates have been synthesized. However, this fascinating phenomenon remains confined only to cuprates, and has not been extended even to neighboring nickelates. At the extreme forefront of research in superconductivity is to explore the possibility of high-temperature superconductivity in non-copper oxides, which will also be quite important for understanding the superconducting mechanism yet unexplained. The common features shared by all high-$T_c$ cuprates are (1) two-dimensional $CuO_2$ planes in crystal structure and (2) $3d^{9\pm\delta}$ configuration in electronic structure. The former feature can be found in other "layered" perovskite ($K_2NiF_4$ structure, etc.) oxides whereas the latter is very rare in ionic solids except for divalent $Cu^{2+}$. The formal similarity between $Ni^{1+}$ and $Cu^{2+}$ seems to suggest that $Ni^{1+}$ compounds might be one possibility. $Ni^{1+}$ is an unusual oxidation state, but Crespin *et al.* reported the monovalent nickel oxide, $LaNiO_2$, in 1983 [1, 2]. $LaNiO_2$ has not only $3d^9$ configuration but also the so-called "infinite-layer" structure, isostructural to $SrCuO_2$, the parent compound of superconducting $Sr_{0.9}La_{0.1}CuO_2$ with $T_c$ = 43 K. Hence this compound might provide a platform for possible high-temperature superconductivity. According to the original report by Crespin *et al.*, $LaNiO_2$ can be synthesized by a special synthetic route, namely a topotactic reduction with $H_2$ of perovskite $LaNiO_3$ at low temperatures (250-450$^o$C). The synthesis by Crespin *et al.*, however, involves complicated steps in a hydrogen recirculating system. In fact, several unsuccessful attempts by other researchers to reproduce their experiments cast some doubt on the existence of the $LaNiO_2$ phase [3]. Later, in 1999, Hayward *et al.* succeeded in reducing $LaNiO_3$ into $LaNiO_2$ with NaH, the most powerful reducing agent, at lower reduction temperatures (~200 $^{o}$C) to avoid the decomposition [4]. Their experiments revealed that the $Ni^{1+}$ ($d^9$) two-dimensional sheets behave quite differently from their isoelectronic $Cu^{2+}$ counterparts, and show no antiferromagnetic long range order. They suggested that this difference may be due to reduced covalent mixing of Ni3*d* and O2*p* orbitals.

$LaNiO_2$ is an interesting compound and may shed some light on the superconducting mechanism of high-$T_c$ cuprates. However, its synthesis is not as easy as for cuprates, and only powder samples are available so far, which prevents detailed measurements of physical properties. In this article, we report the synthesis of epitaxial $LaNiO_2$ thin films by metal organic decomposition (MOD) and subsequent hydrogen reduction. Our synthesis process is much easier than those employed by Crespin *et al*. and Hayward *et al*. This is because of the advantage of thin films with a large-surface-to-volume ratio, which makes oxygen diffusion prompt. Furthermore resistivity measurements were performed on $LaNiO_2$ for the first time, indicating that $LaNiO_2$ is essentially a metal but unfortunately with no trace of superconductivity yet.

**2. Experimental**

We have attempted both vacuum and hydrogen reduction of perovskite $LaNiO_3$ films toward the synthesis of infinite-layer $LaNiO_2$. The starting materials, $LaNiO_3$ films, were prepared by MOD, using La and nickel 2-ethylhexanoate solutions. The stoichiometric mixture of solutions was spin-coated on various substrates, including $SrTiO_3$(100), $LaAlO_3$(100), $DyScO_3$(110), and $YAlO_3$(110) (abbreviated as STO, LAO, DSO, and YAO, respectively) [5]. Table I summarizes the lattice constants of $LaNiO_3$, $LaNiO_2$, and those substrates. The coated films were first calcined at 450°C in air for 10 min to obtain precursors, then fired at 900°C for 3 hours in a tubular furnace under pure $O_2$ ($P(O_2)$ = 1atm). One cycle of spin-coating and calcination gives a film with thickness of ~800 Å after firing. Thicker films were produced by repeating the cycle before final firing. After firing, the films were furnace-cooled in oxygen, and underwent a subsequent reduction process in another tubular furnace. The reduction was made either in vacuum (<$10^{-4}$ torr) or in pure hydrogen ($P(H_2)$ = 1atm) with the reduction temperature ($T_{red}$) and reduction period ($t_{red}$) varied. After reduction, the films were furnace-cooled in the same atmosphere as during reduction. The crystal structure of the films was determined by $\theta$-$2\theta$ scans using a powder X-ray diffractometer, and resistivity was measured by a 4-probe method.

## 3. Results and discussion

### 3.1. Properties of $LaNiO_3$ films grown by MOD

Figures 1(a) shows the XRD patterns around the (200) reflection of the starting $LaNiO_3$ films on different substrates. All the observed peaks between $2\theta$ = 10° and 60° can be indexed as the ($h$00) reflection lines of the perovskite structure, indicating single-crystalline films achieved by solid-state epitaxy. With regard to the substrate dependence, we have observed a noticeable change in the peak positions on different substrates, and also found that the peak intensities of the films are the strongest with best lattice-matched LAO (see table I). A similar trend is also observed in the resistivity. Figures 1(b) shows the temperature dependence of resistivity [$\rho(T)$] for $LaNiO_3$ films on different substrates. The resistivity is the lowest on LAO and the highest on DSO. The resistivity value of $LaNiO_3$ films on LAO is ~100 μΩcm at 300

K and ~10 μΩcm at 4.2 K, which is comparable to the best value reported for bulk samples [6]. Both of the XRD and resistivity data indicated that high-quality epitaxial thin films of $LaNiO_3$ can be obtained by MOD, especially with LAO substrates. Therefore we performed subsequent reduction experiments mostly for films on LAO.

### 3.2. Vacuum reduction

We first attempted vacuum reduction. Figure 2 shows the XRD patterns of the films after vacuum reduction at $T_{red}$ = 400°C with different $t_{red}$ from 13 to 56 hours. The film thickness before reduction was ~800 Å. With $t_{red}$ = 13 hours, the $LaNiO_3$ peaks shift to lower angles, then with $t_{red}$ = 26 hours, the original peaks of $LaNiO_3$ disappear (may be buried under the substrate peaks) and new peaks appear at $2\theta = 11.6^o$ and 35.3°, which arise from double-perovskite $La_2Ni_2O_5$. This suggests that oxygen is gradually removed from the perovskite structure, resulting initially in the formation of oxygen-deficit $LaNiO_{3-\delta}$ with no oxygen order, and then in the formation of $La_2Ni_2O_5$ with the double-perovskite structure. In $La_2Ni_2O_5$, the chains of $NiO_6$ octahedra lying along the *c* axis and $NiO_4$ square-planar units alternate in the *a*-*b* plane [7, 8]. $La_2Ni_2O_5$ films are transparent and insulating. Prolonged reduction at $T_{red}$ = 400°C for $t_{red}$ up to 56 hours did not cause any change in the XRD patterns, indicating no further removal of oxygen from $La_2Ni_2O_5$. Reduction in vacuum at $T_{red}$ higher than 500°C lead to the decomposition of the perovskite phase. These observations concluded that it is impossible to obtain infinite-layer $LaNiO_2$ by vacuum reduction in a practical time scale.

### 3.3. Hydrogen reduction

Next we attempted hydrogen reduction, following the works by Crespin *et al.* [1, 2]. Figures 3(a) shows the XRD patterns after hydrogen reduction at $T_{red}$ = 350$^{o}$C with different $t_{red}$ from 1 to 7 hours. The film thickness before reduction is ~1600 Å. With $t_{red}$ = 1 hour, the $LaNiO_3$ peaks shift to lower angles as seen in vacuum reduction with $t_{red}$ = 13 hours. With $t_{red}$ = 2 hours, new peaks appear in addition to the peaks of oxygen-deficit $LaNiO_{3-\delta}$. One set of peaks give $d$ = 3.969 Å, which agrees with the $a$-axis lattice constant of infinite-layer $LaNiO_2$, and the other set of peaks give $d$ = 3.368 Å, which agrees with the $c$-axis lattice constant of the same compound. The former peaks arise from $a$-axis oriented $LaNiO_2$ grains and the latter from $c$-axis oriented grains. With increasing $t_{red}$ from 2 to 7 hours, the peaks of oxygen-deficit $LaNiO_{3-\delta}$ gradually diminish. At the same time, the peaks of $a$-axis oriented $LaNiO_2$ grains grow although those of $c$-axis oriented grains fade away, indicating the conversion of $c$-axis to $a$-axis grains. Figures 3(b) is the corresponding $\rho(T)$ data. The resistivity is the lowest for $t_{red}$ = 2 hours, and ~4 mΩcm, at best, with almost no temperature dependence, indicating a metallic nature of $LaNiO_2$. This indicates that only $c$-axis oriented grains of infinite-layer $LaNiO_2$ contribute to electron transport. The metallic nature of $LaNiO_2$ is consistent with the LDA (local density approximation) band calculation by Lee and Pickett [9], but inconsistent with the band calculation by Anisimov *et al.* [10], taking account of strong electron correlation (LDA + U), which gave a stable antiferromagnetic insulator very much like that of $Cu^{2+}$ oxides.

We have investigated the substrate and thickness dependence in order to maximize the amount of $c$-axis oriented grains of infinite-layer $LaNiO_2$. The results are summarized in Figs. 4. All the films in this figure were reduced with $T_{red}$ = 350$^{o}$C and $t_{red}$ = 2 hours. The change of a substrate from LAO to STO seems to enhance the peak intensities of *c*-axis grains but with no noticeable change in resistivity [11]. The increase in film thickness on LAO enhances the peak intensities of *c*-axis grains, and lowers the resistivity to some extent. We can explain these substrate and thickness dependences in the stabilization of *c*-axis grains as follows. The in-plane lattice constant expands substantially from 3.817 Å to 3.959 Å by conversion from perovskite $LaNiO_3$ to infinite-layer $LaNiO_2$. It is natural to think that the substrates lattice-matched to infinite-layer $LaNiO_2$ in plane may favor its *c*-axis growth. From this point of view, STO is a fair choice (DSO may be a better choice) whereas LAO not an appropriate choice. Furthermore we can speculate that thinner films on LAO suffer more from the effect of epitaxy, resulting in more difficulty for *c*-axis growth as was actually observed. Our results seem to indicate that it may not be easy to produce *c*-axis oriented single-crystalline $LaNiO_2$ films although it is possible to produce single-phase $LaNiO_2$ by MOD.

**4. Summary**

We report the synthesis of infinite-layer $LaNiO_2$ thin films by a rather simple method, namely metal organic decomposition and subsequent topotactic reduction by hydrogen from $LaNiO_3$ to $LaNiO_2$. Almost single-phase thin films of $LaNiO_2$ can be synthesized but with a mixture of *a*- and *c*- axis orientations. It seems that only *c*-axis oriented grains of infinite-layer $LaNiO_2$ contribute to electron transport. We have found that lattice matching between $LaNiO_2$ and substrates is important for the stabilization of *c*-axis oriented grains. Resistivity measurements indicate that $LaNiO_2$ is essentially a metal but unfortunately with no trace of superconductivity yet.

**Acknowledgements**

The authors thank Dr. T. Kumagai for support and encouragement and they also thank Crystec GmbH, Germany for developing new $\mathit{RE}ScO_3$ substrates.

**References**

[1] M. Crespin, P. Levitz, L. Gatineau, J. Chem. Soc. Faraday Trans. **79** (1983) 1181.

[2] M. Crespin, O. Isnard, F. Dubois, J. Choisnet, P. Odier, J. Solid State Chem. **178** (2005) 1326.

[3] M. J. Martinez-Lope, M. T. Casais, J. A. Alonso, J. Alloys Compd. **277** (1998) 109.

[4] M. A. Hayward, M. A. Green, M. J. Rosseinsky, J. Sloan, J. Am. Chem. Soc. **121** (1999) 8843.

[5] $DyScO_3$ and $YAlO_3$ have the $GdFeO_3$, distorted perovskite, structure. The (110) face of $GdFeO_3$ structure is equivalent to the (100) face of pseudo-perovskite structure.

[6] J. –S. Zhou, J. B. Goodenough, B. Dabrowski, Phys. Rev. B **67** (2003) 020404(R).

[7] M. J. Sayagués, M. Vallet-Regí, A. Caneiro, J.M.González-Calbet, J. Solid State Chem. **110** (1994) 295.

[8] T. Moriga, O. Usaka, I. Nakabayashi, T. Kinouchi, S. Kikkawa, F. Kanamaru, Solid State Ionics **79** (1995) 252

[9] K. –W. Lee, W. E. Pickett, Phys. Rev. B **70** (2004) 165109.

[10] V. I. Anisimov, D. Bukhvalov, T. M. Rice, Phys. Rev. B **59** (1999) 7901.

[11] In the case of STO, however, the peaks of *a*-axis grains are completely buried under the substrate peaks, so the relative portion of *c*-axis to *a*-axis grains is not known.

Table I. Lattice constants of $LaNiO_3$, $LaNiO_2$, and substrates used. The lattice constants of $DyScO_3$, $LaAlO_3$, and $YAlO_3$, which have distorted perovskite structures, are given by regarding these substrates as the pseudo-cubic structure.

| | Lattice constant [Å] |
|---|---|
| $LaNiO_3$ | 3.817 |
| $LaNiO_2$ | a=3.959, c=3.375 |
| $DyScO_3$ | 3.944 |
| $SrTiO_3$ | 3.905 |
| $LaAlO_3$ | 3.790 |
| $YAlO_3$ | 3.715 |

## Figure captions

Figures 1. (a) XRD patterns and (b) $\rho(T)$ curves of $LaNiO_3$ films prepared by metal organic decomposition on different substrates. The inset in (b) is an enlarged view to compare the $\rho(T)$ curves of a best bulk specimen [6] and our MOD film on LAO.

Figure 2. XRD patterns of the films after vacuum reduction at $T_{red}$ = 400$^{o}$C with different reduction periods ($t_{red}$) from 13 to 56 hours.

Figures 3. (a) XRD patterns and (b) $\rho(T)$ curves of films after hydrogen reduction at $T_{red}$ = 350$^{o}$C with different reduction periods ($t_{red}$) from 1 to 7 hours.

Figures 4. Substrate and thickness dependences of (a) XRD patterns and (b) $\rho(T)$ curves of films after hydrogen reduction at $T_{red}$ = 350$^{o}$C for $t_{red}$ = 2 hours. The peak marked by * in the XRD pattern of the film on STO is due to Ag electrodes.

Fig. 1(a). D. Kaneko *et al.*, PCP-3 / ISS2008

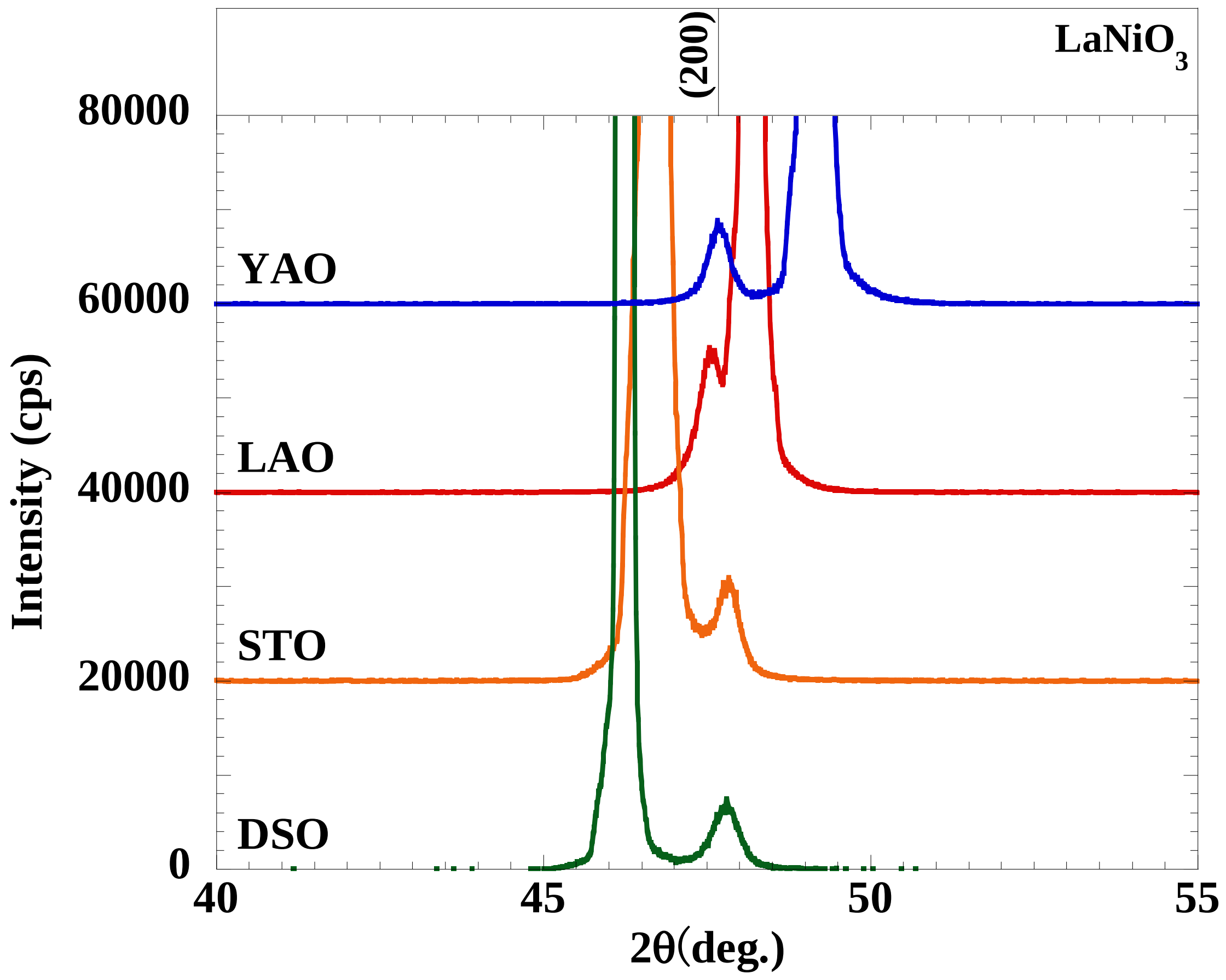

Fig. 1(b). D. Kaneko *et al.*, PCP-3 / ISS2008

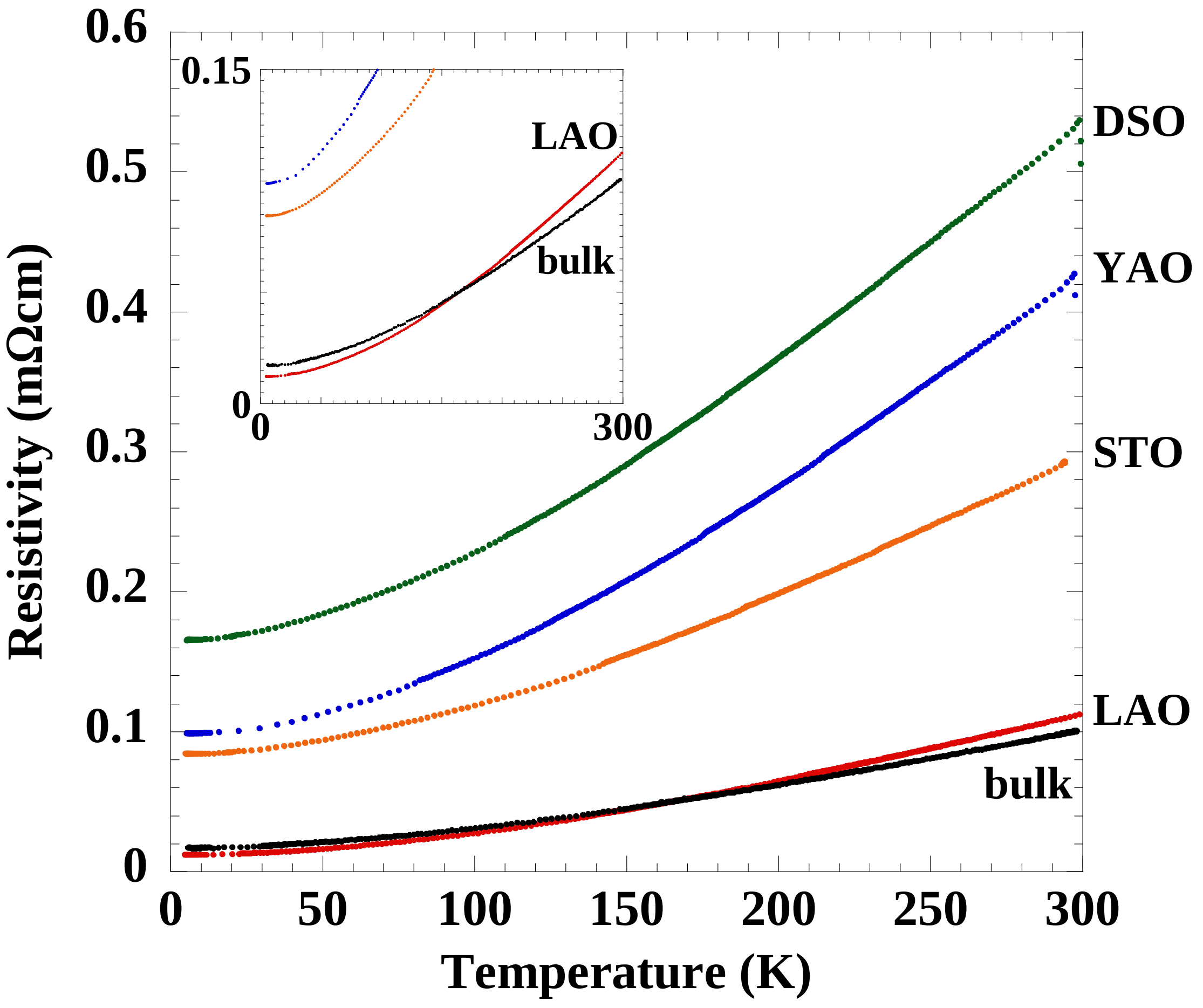

Fig. 2. D. Kaneko *et al.*, PCP-3 / ISS2008

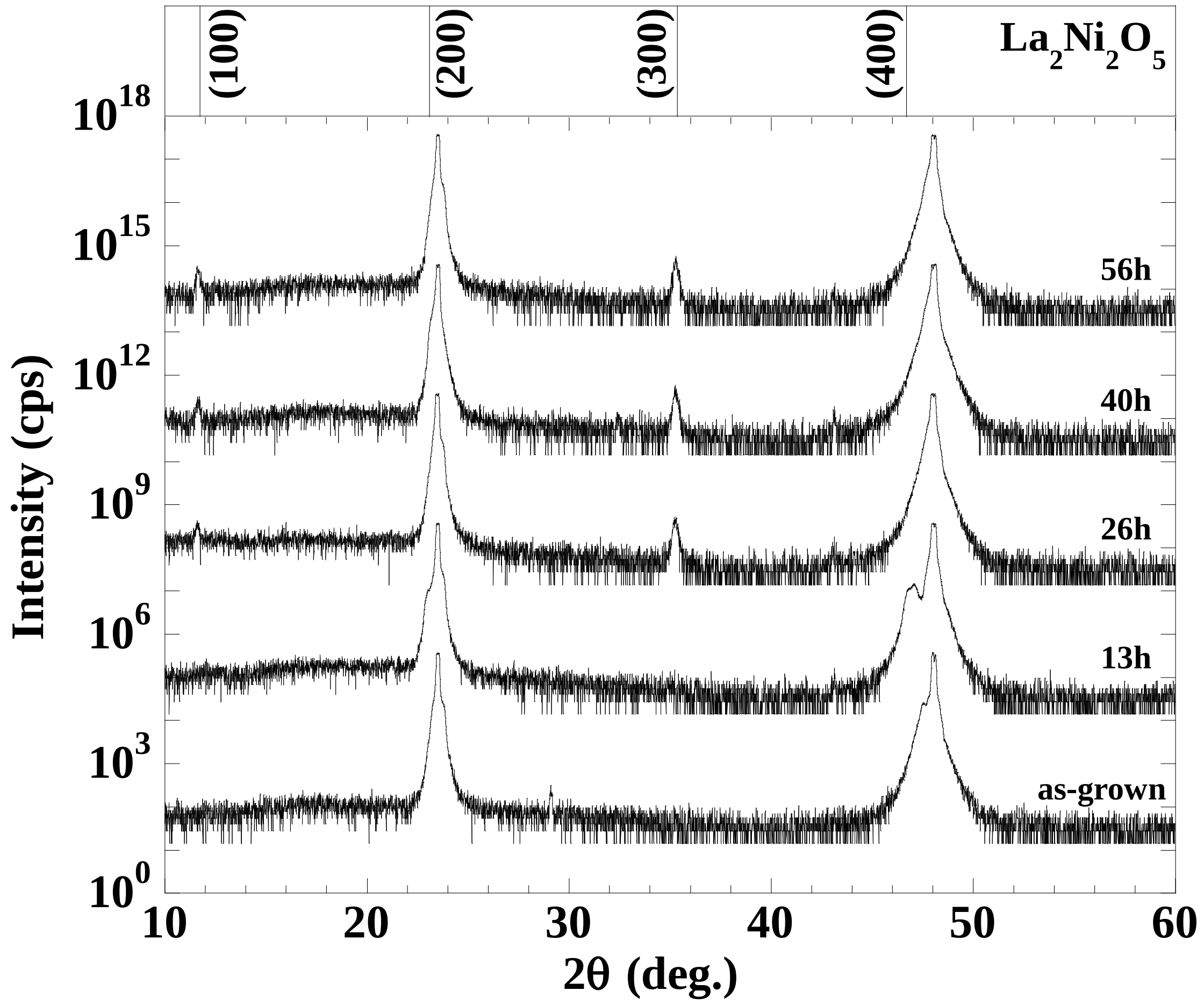

Fig. 3(a). D. Kaneko *et al.*, PCP-3 / ISS2008

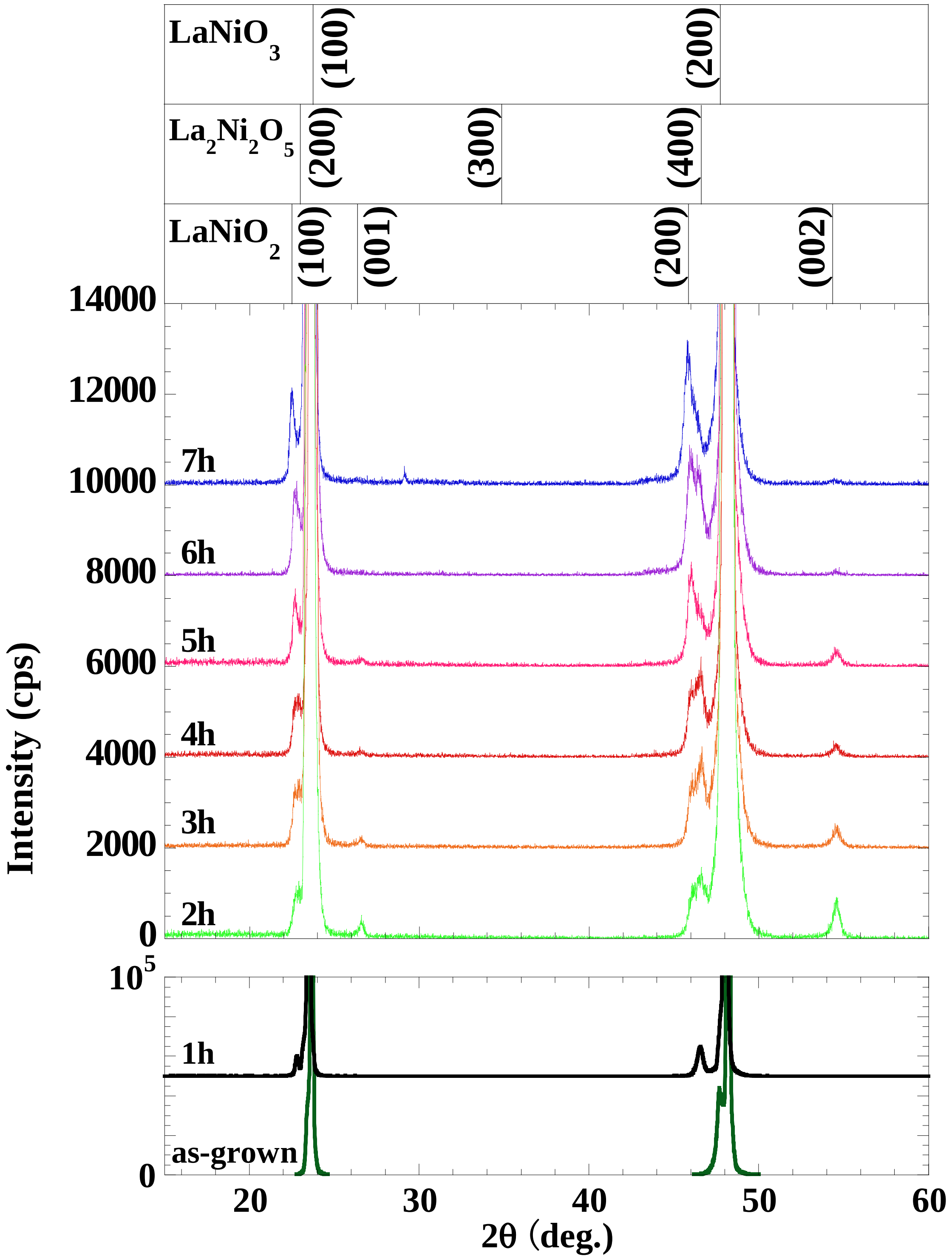

Fig. 3(b). D. Kaneko *et al*., PCP-3 / ISS2008

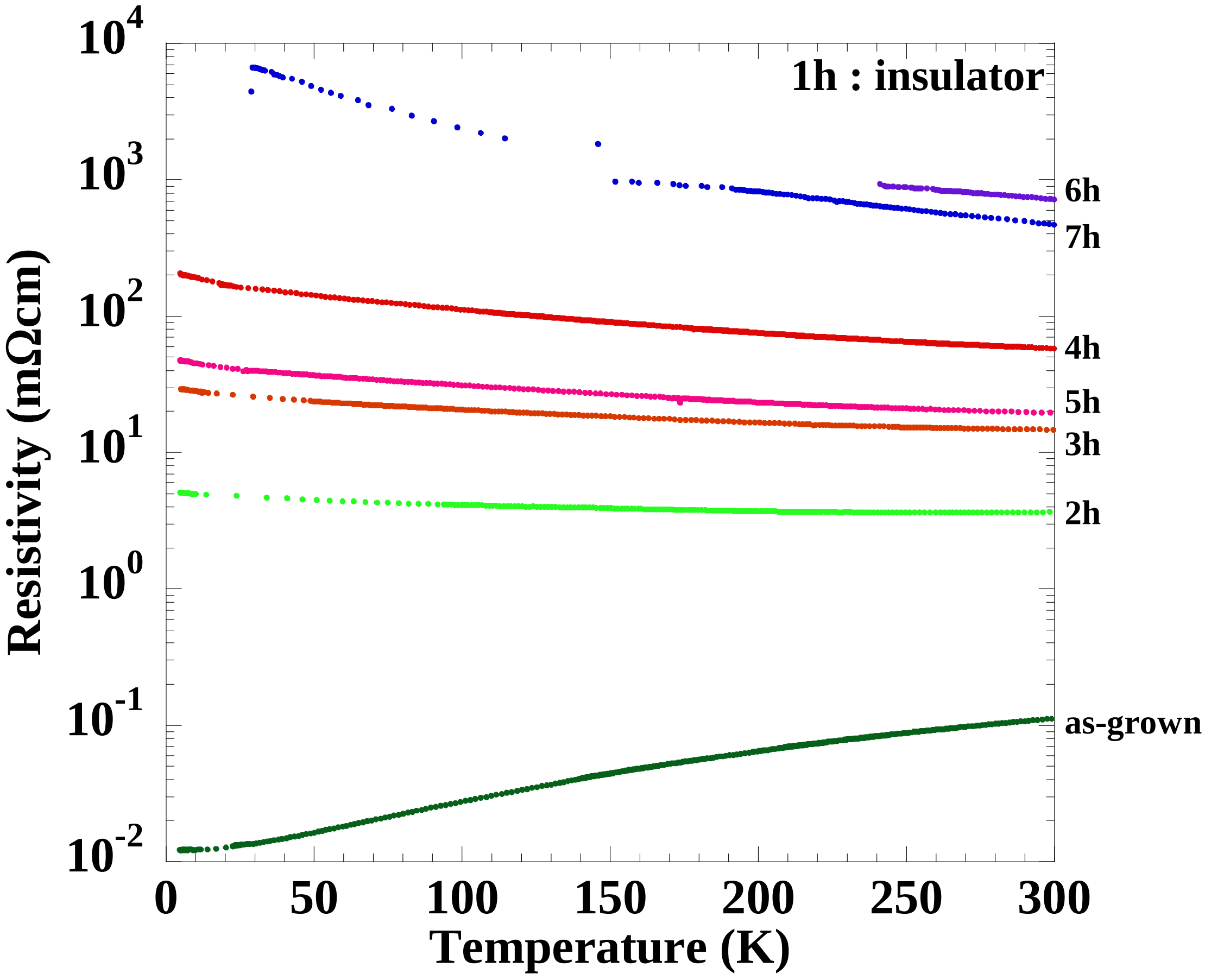

Fig. 4(a). D. Kaneko *et al*., PCP-3 / ISS2008

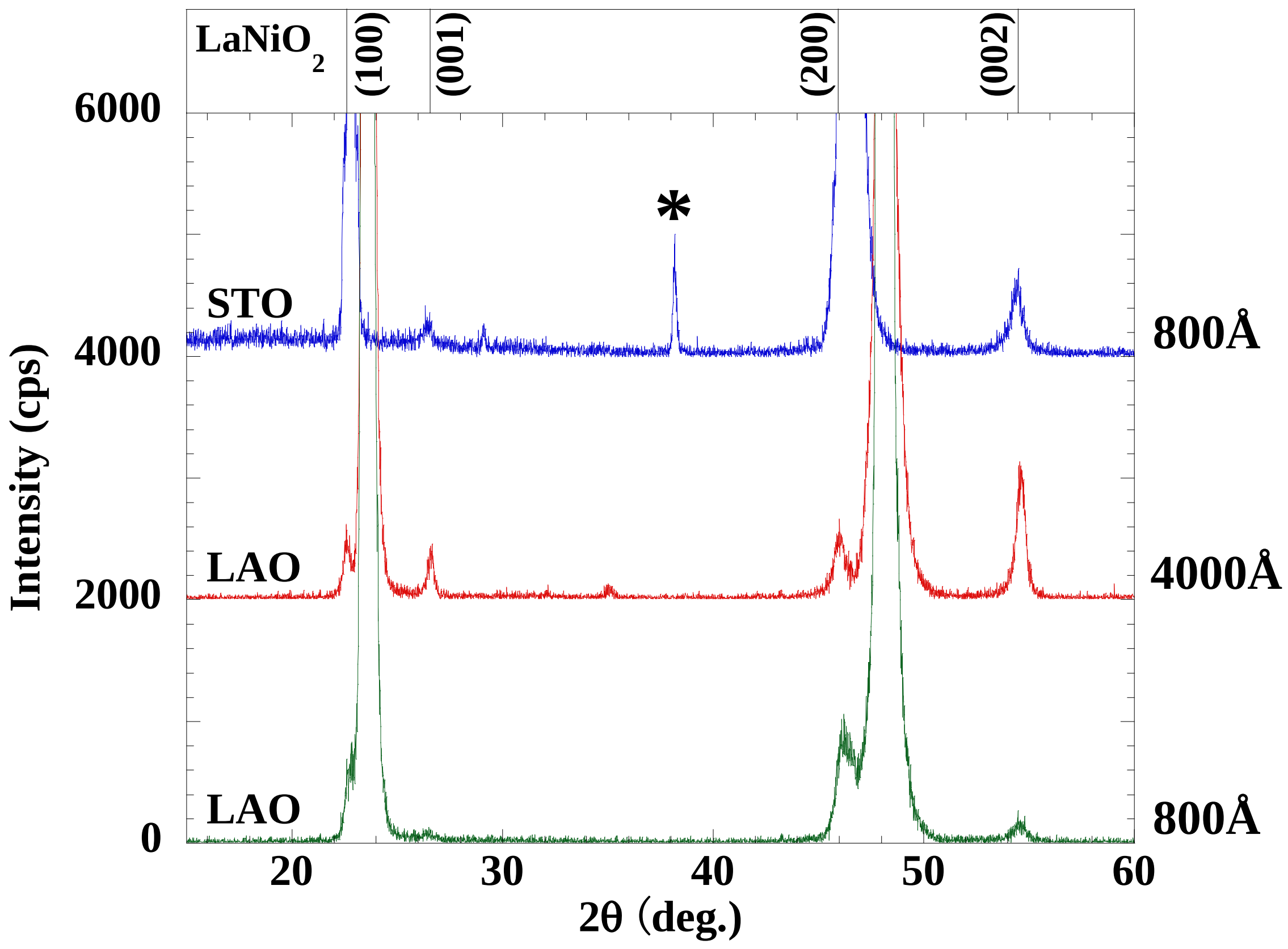

Fig. 4(b). D. Kaneko *et al.*, PCP-3 / ISS2008

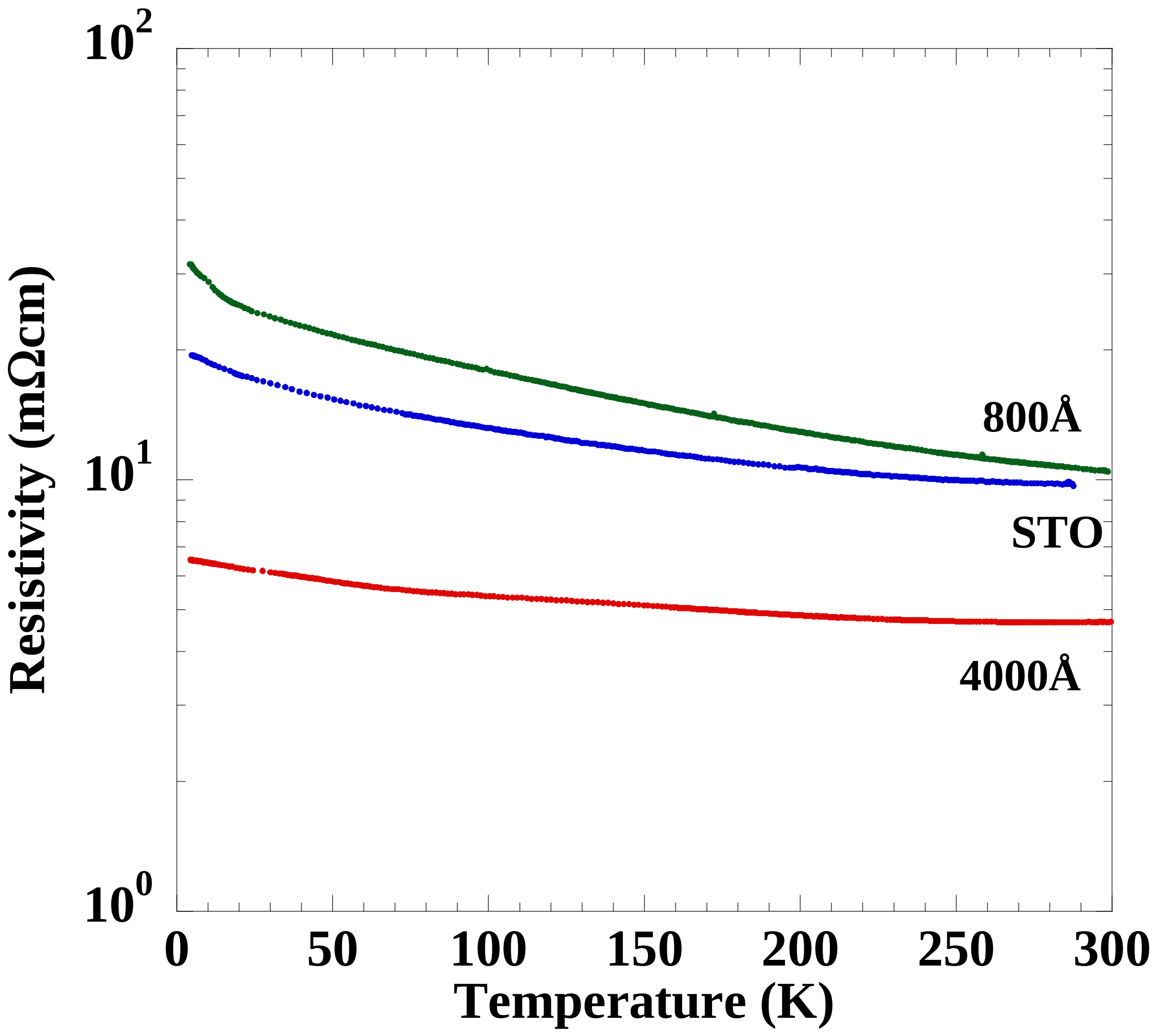